\documentclass[twocolumn,prl,showpacs]{revtex4}
\usepackage{graphicx}
\usepackage{bm}
\begin{document}
\title{The healing mechanism for excited molecules near metallic surfaces}
\author{ B. Barbiellini$^1$ and P.M. Platzman$^2$}

\affiliation{$^1$Physics Department, Northeastern University, Boston MA
02115\\
$^2$Bell Laboratories, Lucent Technology, 600 Mountain Avenue,
Murray Hill, NJ07974}\date{\today}
\begin{abstract}
Radiation damage prevents the ability to obtain
images from individual molecules.
We suggest that 
this problem can be avoided
for organic molecules by placing them
in close proximity with
a metallic surface.
The molecules will then quickly dissipate
any electronic excitation
via their coupling to the
metal surface. They may therefore be observed for
a number of elastic scattering events that is
sufficient to
determine their structure.
\end{abstract}
\maketitle
Examining individual molecules at an atomic length scale
is one of the most challenging modern problems of science.
An analysis of elastic and inelastic cross sections
provides a basis
for understanding why molecules can become
fragmented beyond recognition
long before the time necessary to
accumulate sufficient
counts for image reconstruction.
In the case of organic molecules, most structural knowledge has been gained
using crystalline samples, where damaging exposure
is distributed over many identical molecules,
and the image
is almost the same as that of an undamaged molecule.
For molecules which cannot be crystallized,
there is an urgent need for methods that
can give information on their structures.

In single molecule diffraction techniques proposed using either
femtosecond x-ray pulses \cite{nat_fel}
or high energy electron beams \cite{prl_smd},
short wavelengths
are used to image the atomic structure
of individual molecules, but the radiation doses
destroy the molecule after providing a diffraction
pattern. 
Imaging with coherent electrons from Low
Energy Electron Point Sources
(LEEPS) is a less destructive method to
characterize single molecules
\cite{leeps}.
A holographic interference pattern can be generated by LEEPS 
from which one can reconstruct the object's structure.
For example, one can reconstruct the
structure of single polymer strands like DNA \cite{fink_nat}.
Recently, another method using interference patterns with standing
x-ray waves has been used to map the 
perylene-tetracarboxylic-dianhydride (PTCDA) 
molecule on a silver surface \cite{nixsw05}. 
The use of positrons for microscopy is 
not as well known,
but it has tremendous potential.
Positrons, owing to their positive charge,
have negative work functions in several materials,
and they can be reemitted spontaneously.
This feature allows the control of slow positron
beams having energies in the epithermal region.
A positron microscope for imaging
single molecules has been recently proposed
by Mills and Platzman \cite{mills01}.
In order to get proper structural information
with sufficient resolution,
Mills and Platzman have estimated
that each atom in a molecule of
$200$ atoms should suffer about $50$
annihilations in the course of
the positron diffraction.
After the annihilation,
the molecule is
in an ionic excited state, and
under normal circumstances, 
such a large number of
ionizations would destroy the molecule.

Clearly, while exploring the
opportunities of imaging single molecules
with either x-rays, electrons
or positrons, it is vital to reduce
radiation damage at the
level of individual molecules.
In this letter, 
we utilize the neutralization process when a molecule
is placed near a metallic surface - here called the 
{\em healing mechanism} - and propose that it may be applied to 
heal the radiation damage.
The main idea is to fill
the vacant molecular orbital
by an electron from the metal
in a time short compared to some
characteristic vibration time of the molecule.
Therefore, charge neutrality
is restored before the onset of the vibrational modes
which would lead to the destruction of the molecule.
This healing mechanism 
can be decomposed in two main steps. 
In the first step, when the positive
ion is sufficiently close to the surface,
one conduction electron of the metal will
tunnel under the influence of the strong
electric field produced by the ion.
In the second step, the tunneling electron will
fill the hole via an Auger process involving an
electron-hole pair of the metal.
The physics of the healing mechanism is closely
related to various neutralization processes 
\cite{relax0,relax1,relax2,relax3,relax4,relax5,relax6,relax7}
and to the intermolecular Coulombic decay \cite{icd}.
Most of this work describes collisions of slow ions 
with solids.  Our study involves the neutralization
of physisorbed molecules on metallic surfaces.
Given the wide range of
estimates of the neutralization time 
in previous work \cite{relax1},
we must perform a conservative
estimate to demonstrate the feasibility of the 
healing mechanism in our case.

\begin{figure}
\begin{center}
\includegraphics[width=\hsize]{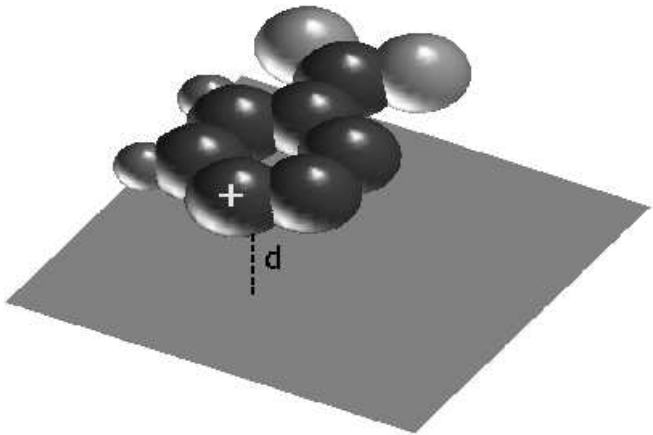}
\end{center}
\caption{
Schematic of a typical healing mechanism set-up:
an ionized molecule is placed above a metallic substrate
and the positive charge is at a distance $d$ from the
substrate. In this example, atoms of a PTCDA fragment
are shown. The small grey spheres denote H, the large
grey spheres denote O and the dark large spheres denote C. 
}
\label{setup}
\end{figure}

A schematic of the healing mechanism set-up is
shown in Fig.~\ref{setup}.
The molecule, in this example PTCDA, 
is on a metallic substrate,
which should be chemically inert 
to preserve the integrity of the molecule. Gold, silver or 
other noble metals are possible substrates.
In reality, the PTCDA geometry might slightly bend 
near the metallic substrate because of the readiness 
of some PTCDA atoms to form
weak bonds with the surface \cite{nixsw05}.

When a positive charge is suddenly created on the molecule,
the resulting electric field attracts the electrons
of the metal. 
The probability $T$ that one electron 
escapes from
the metal depends on the work function $\Phi$ 
and on the distance $d$ of the positive charge
from the surface.
An accurate determination of the tunneling probability
$T$ would require solving the Sch\"odinger equation, 
for example
using the matching wave function method
\cite{fowler}. However, the result
is mostly dominated by an exponential
factor given by the WBK approximation
\cite{probst}
\begin{equation}
T=\exp(-2\int_{0}^{s}dz~\sqrt{2m(V(z)-E_F)}/\hbar)~,
\end{equation}
where $s$ is the width of the potential barrier 
and $E_F$ is the Fermi energy of the metal.
The actual shape of the potential barrier
$V(z)$ is complicated, but
a linear approximation of
the electrostatic potential 
at the metal-molecule
interface yields a simple 
triangular barrier, and
the penetration probability 
$T$ becomes \cite{footnote1}
\begin{equation}
T(d)=\exp(-\sqrt{2}~
\frac{2}{3}\Phi^{3/2}d^2)~,
\label{EqT}
\end{equation}
where atomic units have been used.
\begin{figure}
\begin{center}
\includegraphics[width=\hsize]{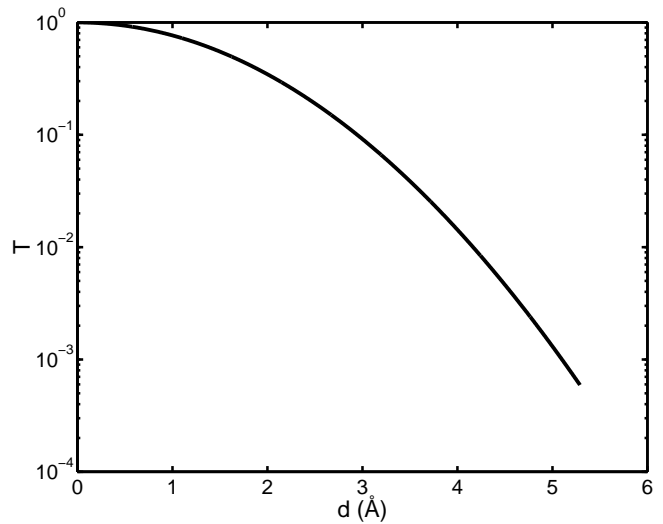}
\end{center}
\caption{
Penetration probability $T$
as a function of the distance $d$
for a work function $\Phi=5$ eV.
}
\label{fig1}
\end{figure}
Fig.~\ref{fig1} shows that $T$ 
under a barrier characterized
by a work function $\Phi=5$ eV
is still about $0.01$
for a distance $d=4~\mbox{\AA~}$.
Actually, our major assumption - the triangular barrier -
underestimates $T$ since the electric field gets
stronger as an electron approaches
the molecule. Our assumption, as we shall see,
therefore leads to an overestimate
of the healing time.

\begin{figure}
\begin{center}
\includegraphics[width=\hsize]{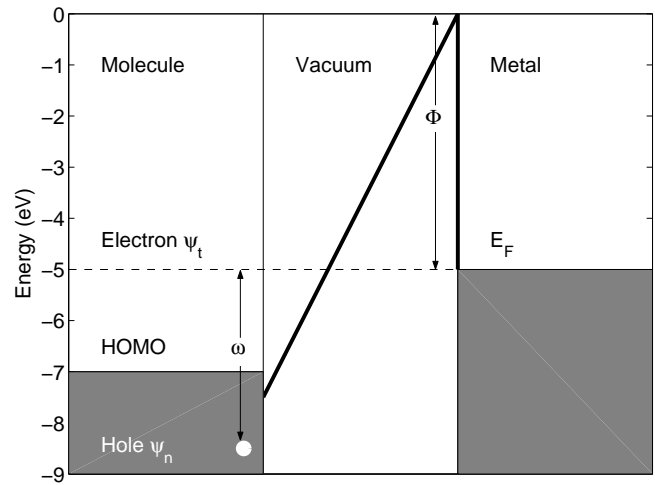}
\end{center}
\caption{
Schematic diagram of the energy levels
at the organic metal-molecule junction.}
\label{diag}
\end{figure}

An electron of the metal
is expected to fall into the vacant orbital
of the molecule.
Indeed, a photoemission study \cite{kahn96} of 
the PTCDA organic molecule on a metal surface
reveals the energy-level alignment shown
in the schematic diagram of Fig.~\ref{diag},
where the highest occupied molecular 
orbital (HOMO) level is about $2$ eV
below $E_F$.
A similar diagram of energy levels has been
obtained via first-principles 
calculations \cite{picozzi}.

The energy separation $\omega$ (see Fig.~\ref{diag}) 
between the state $\psi_t$ of the electron spilling 
out from the metal \cite{footnote_ef}
and the molecular orbital of the hole
$\psi_n$ must be dissipated during the recombination.
An Auger process involving an electron-hole 
excitation in the metal can mediate the capture
of the electron in the state $\psi_t$
into the vacant molecular state $\psi_n$.
There are several possible 
electron transfer processes between molecules 
and surfaces \cite{surfsci}.
For chemisorbed systems, a hole lifetime is much
shorter than a typical vibration time \cite{gunnarsson}, 
so that normally the hole would
be filled before the molecule can fly apart.
For N$_2$ molecules physisorbed on graphite, 
experiments suggest that
the neutralization
occurs with a characteristic 
time $\tau\approx 10^{-14}$ s \cite{nilsson}.
In general, for a molecule far from
the surface, the corresponding
hole may not be filled for a while. 
In this case, the relevant process 
is the Auger neutralization
as described by Propst \cite{probst}. 
The actual calculation of the Auger 
neutralization rate 
$1/\tau(d,\omega)$
is a difficult problem
including subtle screening effects at the 
metal-vacuum interface \cite{relax1}.
Nevertheless, it has been suggested by
numerical calculations that, as 
a good approximation,  
the dependence on the distance $d$
and the energy $\omega$ factorize
\cite{relax2} as
\begin{equation}
\frac{1}{\tau}(d,\omega)=f(d)\frac{1}{\tau_B}(\omega),
\label{eq_factor}
\end{equation}
where $1/\tau_B(\omega)$ is obtained using
the formula for the molecule immersed in the bulk 
and $f(d)$ is an unknown function, 
usually approximated 
by an exponential \cite{relax2}, 
which decays away from the 
metal surface.
Propst \cite{probst}, however, 
has shown that the distance of the
molecule from the surface comes into play
mostly through the WKB transmission
probability. Therefore, we can assume $f(d)\approx T(d)$.
The matrix element $M_{\alpha}({\bf q})$ of the Auger process
in the bulk involves the ground state $u_0$
of the metal, the excited state $u_{\alpha}$ of
the metal and the density
operator $\rho^+_{\bf q}$
for excitations with momentum transfer ${\bf q}$.
The actual form of $M_{\alpha}({\bf q})$ \cite{bock} reads 
\begin{equation}
M_{\alpha}({\bf q})= F({\bf q})V_C(q)<u_{\alpha}|\rho^+_{\bf q}|u_0>,
\end{equation}
where
$F({\bf q})$ is a form factor given by
\begin{equation}
F({\bf q})=\int d{\bf r}~\psi_n({\bf r}) 
\psi_t({\bf r}) \exp(i{\bf q r})~,
\end{equation}
$V_C(q)$ is the Coulomb interaction Fourier transform
\begin{equation}
V_C(q)=\frac{4\pi e^2}{q^2L^3}~,
\end{equation}
and $L^3$ is the volume of the metallic sample.
The rate of the Auger recombination can be calculated
by using Fermi's golden rule
\begin{equation}
\frac{1}{\tau_B}=\sum_{\bf q}\sum_{\alpha} |M_{\alpha}({\bf q})|^2
\delta(\omega-\omega_{\alpha0})~,
\label{eq_Fermi}
\end{equation}
where $\omega_{\alpha0}$ is the
energy difference between the metal ground-state
$u_0$ and the excited state $u_{\alpha}$.
Eq.~\ref{eq_Fermi} can be rewritten in the form 
\begin{equation}
\frac{1}{\tau_B}=\sum_{\bf q} S({\bf q},\omega) 
V_c({\bf q})^2 F({\bf q})^2 ,
\label{eq_as}
\end{equation}
where $S({\bf q},\omega)$ is 
the dynamical structure factor \cite{pines} 
defined by
\begin{equation}
S({\bf q},\omega)=\sum_{\alpha} |<u_{\alpha}|\rho^+_{\bf q}|u_0>|^2
\delta(\omega-\omega_{\alpha0})~.
\end{equation}
The fluctuation-dissipation theorem relates
the dynamical structure factor $S({\bf q},\omega)$ to
the dielectric function
$\epsilon({\bf q},\omega)$ by the formula
\begin{equation}
S({\bf q},\omega)=
-\frac{1}{V_C(q)}\mbox{Im}\left
[ \frac{1}{\epsilon({\bf q},\omega)} 
\right ]~.
\label{eq_diss}
\end{equation}
Substituting Eq.~\ref{eq_diss} in Eq.~\ref{eq_as},
we obtain
\begin{equation}
\frac{1}{\tau_B}=-\int \frac{L^3d^3{\bf q}}{8\pi^3}
~\mbox{Im}\left [ \frac{F({\bf q})^2V_C(q)}
{\epsilon({\bf q},\omega)} \right ]~.
\label{Eq_Auger}
\end{equation}
To extract an Auger rate estimate, 
one can use the Lindhard
RPA formula \cite{pines} for
the dielectric function
$\epsilon(q,\omega)$.
Density inhomogeneity effects,
due to the lattice and the surface,
and correlation corrections beyond 
the RPA 
are not considered in this study.
For $\psi_n(r)$, one can use
a Slater type of orbital \cite{slater} (STO) given by
\begin{equation}
\psi_n(r)=\sqrt{\frac{\zeta^3}{\pi}}\exp(-\zeta r)~,
\end{equation}
where $1/\zeta$ provides the size of the molecular orbital.
Therefore, assuming that $\psi_t({\bf r})$ is slowly varying,
the main form factor contribution to the integral
in Eq.~\ref{Eq_Auger} is given by
\begin{equation}
F(q)=\sqrt{\frac{8\zeta^3}{\pi^2}}
\frac{\zeta^{5/2}}{(\zeta^2+q^2)^2}~.
\label{eq_fh}
\end{equation}
\begin{figure}
\begin{center}
\includegraphics[width=\hsize]{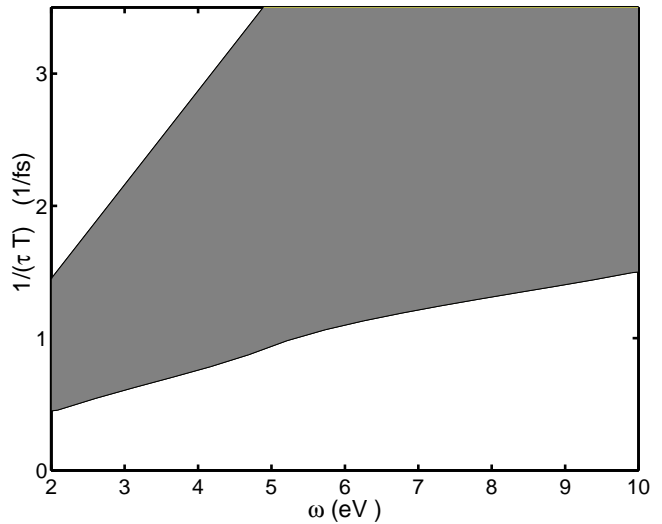}
\end{center}
\caption{
Auger rates in the energy range $2-10$ eV
for a metal with density parameter $r_s=3.1$, 
as in gold. The grey region corresponds to 
$\zeta$ 
ranging from $1$ to $\infty$.}
\label{fig2}
\end{figure}
Fig.~\ref{fig2} shows typical $1/\tau$ values
in the energy range $2-10$ eV
for a gold surface (corresponding 
to a electron density parameter $r_s=3.1$).
Since this rate is proportional to
the penetration
probability $T$, we show
$T^{-1}/\tau$ in the ordinate axis
of the plot in Fig~\ref{fig2}.
The grey region in Fig.~\ref{fig2}
corresponds to a STO exponent $\zeta$
ranging from $1$ to $\infty$.
For a distance of $\sim 4~\mbox{\AA}$,
since $T$ is about $0.01$
the Auger rate becomes
of the order of $10^{13}$ 1/s
in the lower part of the grey region 
in Fig.~\ref{fig2}.
This value is comparable
to the highest vibration modes
in molecules.
All this implies that,
within distances of less than
$4~\mbox{\AA}$,
the time to fill the hole
is estimated to be fast enough
to keep the molecule from being
damaged by molecular vibrations.
As mentioned previously, 
the value of the penetration probability $T$
given by Eq.~\ref{EqT} is a lower bound.
This fact adds to the robustness of the present
model.

In conclusion, the effects of radiation damage on molecules
are of concern, because they can be a
major obstacle to the emergence of microscopy
as a tool to determine the structure of delicate biological 
molecules.
This is true whether x-rays, electrons or positrons are used
as the illumination.
We have shown that in favorable conditions,
ionized molecules on a metal surface can quickly
dissipate any electronic
excitation via their coupling to
the metal.
The relaxation is achieved via an electron tunneling from the metal
to the molecule and falling into the vacant orbital.
The falling electron gives up its
energy to
an excitation of the metal produced by an Auger process.
If the molecule is within distances
of about $4~\mbox{\AA}$ from the metal
surface,
short lifetimes of the hole on the molecule 
(of the order of $10-100$ fs) prevent the
destruction of the molecule.
Our model may in fact overstimate the healing time.
This healing mechanism is particularly suitable
for the positron microscope newly proposed
by Mills and Platzman \cite{mills01},
but it can be applied also to other microscopes
imaging single molecules.
In all these cases,
the molecule on the metallic surface may be observed
without being
damaged for a number of scattering
events sufficient to determine its structure
by speckle diffraction.
These measurements would be ideal for studying
delicate biological molecules
which cannot be crystallized.

We acknowledge Rolando Saniz, Olle Gunnarsson and Dan Nissenbaum for 
useful discussions. This work is supported by the US Department of Energy 
contract DE-AC03-76SF00098 and benefited from the allocation of 
computer time at NERSC and Northeastern University's Advanced 
Scientific Computation Center (ASCC).


\begin{thebibliography}{99}
\bibitem{nat_fel}
R. Neutze {\em et al.},
Nature {\bf 406}, 752 (2000).
\bibitem{prl_smd}
J. C. H. Spence and R. B. Doak,
Phys. Rev. Lett. {\bf 92}, 198102 (2004).
\bibitem{leeps}
H. W. Fink, W. Stocker, and H. Schmid,
Phys. Rev. Lett. {\bf 65}, 1204 (1990).
\bibitem{fink_nat}
H. W. Fink and C. Sch\"onenberger, 
Nature {\bf 398}, 407 (1999).
\bibitem{nixsw05}
A. Hauschild {\em et al.}, 
Phys. Rev. Lett. {\bf 94}, 036106 (2005).
\bibitem{mills01}
A. P. Mills and P. M. Platzman in
{\em New Directions in Antimatter Chemistry and Physics},
C. M. Surko and F. A. Gianturco, eds.,
Kluwer Academic Publishers, The Netherlands, (2001).
\bibitem{relax0}
R.A. Baragiola and C. A. Dukes,
Phys. Rev. Lett. {\bf 76}, 2547 (1996). 
\bibitem{relax1}
M.A. Cazalilla {\em et al.},
Phys. Rev. B {\bf 58}, 13991 (1998).
\bibitem{relax2}
Y. Bandurin, V. A. Esaulov, L. Guillemot, 
and R. C. Monreal, 
Phys. Rev. Lett. {\bf 92}, 017601 (2004).
\bibitem{relax3} 
J. Burgdorfer {\em et al.}, Nucl. Instr. and
Meth. B {\bf 205}, 690 (2003).
\bibitem{relax4}
L. Wirtz {\em et al.}, Phys. Rev. A
{\bf 67}, 012903 (2003).
\bibitem{relax5}
J. St\"ockl {\em et al.},
Phys. Rev. Lett. {\bf 93}, 263201 (2004).
\bibitem{relax6}
T. Fonden and A. Zwartkruis, Phys. Rev. B {\bf 48}, 
15 603 (1993). 
\bibitem{relax7}
S. Wethekam, A. Mertens and H. Winter, 
Phys. Rev. Lett. {\bf 90}, 037602 (2003).
\bibitem{icd} 
R. Santra and L.S. Cederbaum,
Phys. Rep. {\bf 368}, 1 (2002). 
\bibitem{fowler}
See e.g. J. W. Gadzuk and E. W. Plummer,
Rev. Mod. Phys. {\bf 45}, 487 (1973).
\bibitem{probst}
F.M. Propst, Phys. Rev. {\bf 129}, 7 (1963).
\bibitem{footnote1}
We have assumed that the charge distribution 
on the molecule is spherical
and we have used the {\em method of images} 
to determine
the electric field. Near the metal-vacuum interface, 
along the direction from the metal surface to the
positive charge, the electric field intensity is 
$F=2e^2/d^2$; see e.g.
L. Landau and E. M. Lifchitz, 
{\em Electrodynamics of Continuous Media},
Pergamon, New York, (1960).
\bibitem{kahn96}
Y. Hirose {\em et al.},
Phys. Rev. B {\bf 54}, 13748 (1996). 
\bibitem{picozzi}
S. Picozzi {\em et al.}, 
Phys. Rev. B {\bf 68}, 195309 (2003).
\bibitem{footnote_ef}
We have assumed that 
$\psi_t$ is at the Fermi level $E_F$.
\bibitem{surfsci} 
See e.g. W. Sesselmann {\em et al.}, Surf. Sci.
{\bf 146}, 17 (1984) and D. Lovric {\em et al.}, 
Surf. Sci. {\bf 189/190}, 
59 (1987).
\bibitem{gunnarsson}
O. Gunnarsson, private communication.
\bibitem{nilsson}
O. Bj\"orneholm {\em et al.}, Phys. Rev. Lett. {\bf 68}, 1892 (1991).
\bibitem{bock}
Related lifetimes have been used in 
the context of electron relaxation in quantum dots, 
see e.g. U. Bockelmann and T. Egeler, 
Phys. Rev. B {\bf 46}, R15574 (1992).
\bibitem{pines}
D. Pines, {\em Elementary Excitations in Solids}, 
W.A. Benjamin Inc,
New York, Amsterdam (1963).
\bibitem{slater}
J.C. Slater, Phys. Rev. {\bf 36}, 57 (1930).
\end{thebibliography}
\end{document}